\documentclass[12pt]{article}

\usepackage[margin=0.95in]{geometry}
\usepackage{graphicx}
\usepackage{subcaption}
\usepackage{amsmath}
\usepackage{dsfont}
\usepackage{amssymb}
\usepackage{abstract}
\usepackage[auth-sc,affil-sl]{authblk}
\usepackage{hyperref}
\usepackage{parskip}
\usepackage{enumerate}
\usepackage{fancyhdr}
\usepackage[xcdraw,table]{xcolor}
\usepackage[round]{natbib}
\usepackage{multirow}
\usepackage{algorithm}
\usepackage{algorithmicx}
\usepackage{algpseudocode}
\usepackage[table]{xcolor}
\usepackage{multirow}
\usepackage{amsthm}
\usepackage[normalem]{ulem}

\usepackage{fancyvrb}

\newcounter{probnum}
\allowdisplaybreaks

\definecolor{tabblue}{rgb}{.870588,.905882,.94902}
\definecolor{gray}{rgb}{0.5,0.5,0.5}
\definecolor{black}{rgb}{0,0,0}
\definecolor{white}{rgb}{1,1,1}
\definecolor{blue}{rgb}{0.0,0.0,1}

\definecolor{green}{rgb}{0,0.5,0}

\definecolor{yellow}{rgb}{1,0.549,0}

\definecolor{red}{rgb}{0.6,0.0,0.0}

\definecolor{darkred}{rgb}{0.9,0.4,0}

\definecolor{purple}{rgb}{0.58,0,0.827}

\definecolor{backgcode}{rgb}{0.97,0.97,0.8}
\definecolor{Brown}{cmyk}{0,0.81,1,0.60}
\definecolor{OliveGreen}{cmyk}{0.64,0,0.95,0.40}
\definecolor{CadetBlue}{cmyk}{0.62,0.57,0.23,0}

\newcommand{\qu}[1]{``{#1}''}
\newcommand{\bv}[1]{\boldsymbol{#1}}

\newcommand{\allocspace}{\mathbb{W}}

\newcommand{\betaThat}{\hat{\beta}}

\newcommand{\ybar}{\bar{y}}
\newcommand{\Ybar}{\bar{Y}}
\newcommand{\YbarT}{\Ybar_T}
\newcommand{\YbarC}{\Ybar_C}
\newcommand{\ybarT}{\ybar_T}
\newcommand{\ybarC}{\ybar_C}

\newcommand{\xbarT}{\bar{\x}_T}
\newcommand{\xbarC}{\bar{\x}_C}

\newcommand{\half}{\frac{1}{2}}

\newcommand{\Y}{\bv{Y}}

\newcommand{\x}{\bv{x}}
\newcommand{\w}{\bv{w}}
\newcommand{\onevec}{\bv{1}}

\newcommand{\y}{\bv{y}}

\newcommand{\reals}{\mathbb{R}}

\newcommand{\beqn}{\vspace{-0.25cm}\begin{eqnarray*}}
\newcommand{\eeqn}{\end{eqnarray*}}
\newcommand{\bneqn}{\vspace{-0.25cm}\begin{eqnarray}}
\newcommand{\eneqn}{\end{eqnarray}}
\newcommand{\benum}{\begin{enumerate}}
\newcommand{\eenum}{\end{enumerate}}

\newcommand{\parens}[1]{\left(#1\right)}

\newcommand{\bracks}[1]{\left[#1\right]}
\newcommand{\braces}[1]{\left\{#1\right\}}

\newcommand{\abss}[1]{\left|#1\right|}

\newcommand{\cexpesubnostr}[3]{\expesubnostr{#1}{#2\,|\,#3}}

\newcommand{\expesubnostr}[2]{\mathbb{E}_{\,#1}[#2]}

\newcommand{\oneover}[1]{\frac{1}{#1}}

\newcommand{\normnot}[2]{\mathcal{N}\parens{#1,\,#2}}

\newcommand{\uniform}[2]{\mathrm{U}\parens{#1,\,#2}}
\newcommand{\stduniform}{\uniform{0}{1}}
\newcommand{\exponential}[1]{\mathrm{Exp}\parens{#1}}

\newcommand{\betaT}{\beta_T}
\newcommand{\errorrv}{\mathcal{E}}
\newcommand{\berrorrv}{\bv{\errorrv}}

\theoremstyle{remark}

\usepackage{xcolor}

\title{Better Experimental Design by Hybridizing Binary Matching with Imbalance Optimization}

\author[1]{Abba M. Krieger\thanks{Electronic address: \texttt{krieger@wharton.upenn.edu}; Prinicipal Corresponding author}}
\author[2]{David Azriel\thanks{Electronic address: \texttt{davidazr@ie.technion.ac.il}; Corresponding author}}
\author[3]{Adam Kapelner\thanks{Electronic address: \texttt{kapelner@qc.cuny.edu};  Corresponding author}}

\affil[1]{\small Department of Statistics, The Wharton School of the University of Pennsylvania}
\affil[2]{Faculty of Industrial Engineering and Management, The Technion, Haifa, Israel}
\affil[3]{Department of Mathematics, Queens College, The City  University of New York}

\begin{document}
\maketitle

\begin{abstract}
We present a new experimental design procedure that divides a set of experimental units into two groups in order to minimize error in estimating an additive treatment effect. One concern is minimizing error at the experimental design stage is large covariate imbalance between the two groups. Another concern is robustness of design to misspecification in response models.  We address both concerns in our proposed design: we first place subjects into pairs using optimal nonbipartite matching, making our estimator robust to complicated non-linear response models. Our innovation is to keep the matched pairs extant, take differences of the covariate values within each matched pair and then we use the greedy switching heuristic of Krieger et al. (2019) or rerandomization on these differences. This latter step greatly reduce covariate imbalance to the rate $O_p(n^{-4})$ in the case of one covariate that are uniformly distributed. This rate benefits from the greedy switching heuristic which is $O_p(n^{-3})$ and the rate of matching which is $O_p(n^{-1})$. Further, our resultant designs are shown to be as random as matching which is robust to unobserved covariates. When compared to previous designs, our approach exhibits significant improvement in the mean squared error of the treatment effect estimator when the response model is nonlinear and performs at least as well when it the response model is linear. Our design procedure is found as a method in the open source \texttt{R} package available on \texttt{CRAN} called \texttt{GreedyExperimentalDesign}.
\end{abstract}
\vspace{5cm}
\pagebreak

\section{Introduction}\label{sec:intro}

The setting we consider is the two-arm non-sequential randomized study with a continuous endpoint (e.g. a pill-placebo double-blind clinical trial assessing blood glucose improvement) that seeks inference for an additive treatment effect. There are $2n$ individuals which are sample-size balanced between the two groups. Each subject $i$ is placed into the treatment or control group and this information is encoded in $w_i \in \braces{-1, +1}$ where -1 denotes assignment to control and +1 denotes assignment to treatment. The vector $\w := \bracks{w_1~\ldots~w_{2n}}$ is called an \textit{assignment}, \textit{allocation}, or a \textit{randomization}. The collection of the legal $\w$ vectors and the probability distribution on the legal vectors in the experimental setting is known as an experimental \textit{design}, \textit{strategy}, \textit{algorithm} or \textit{procedure}.

There are $p$ covariates with the $j$th's value for subject $i$ denoted $x_{ij}$ that are assumed known in advance of the randomization (the setting of known measurements is sometimes called \textit{offline} to distinguish it from the \textit{sequential} setting, the latter not being the setting considered in this paper). After any randomization, the values of the subjects' covariates in each group are approximately the same; and it is this fact that is the main reason randomization is employed when seeking causal inference \citep{Cornfield1959}. Thus, \textit{imbalance} (of which there are many principled metrics) in the covariate values between the treatment subjects and the control subjects should be small. Upon completion, the sample responses $y_i$, the result of an unknown process we call the \emph{response model}, are assessed. These sample responses are used to compute a causal estimate of the population average treatment effect $\beta_T$.

Our contribution is a new experimental design that provides (1) lower error in the measurement of the average treatment effect and (2) is robust when the response model is linear and/or non-linear. If the response model is linear, the best experimental design is one that optimizes the treatment assignment for minimal covariate imbalance which we call \emph{imbalance-optimizing designs}. If the response model is nonlinear, a good strategy is to optimize the treatment assignment via creating binary matches (i.e. nonbipartite pairings of subjects) with small intramatch covariate distance. If the response model is a combination of both linear and nonlinear, then both criteria (minimize covariate imbalance and matching) become important. Our hybrid design does both: first by creating optimal nonbipartite matches on covariate distance and then optimizing intramatch assignments to provide small covariate imbalance across the whole sample. The latter can be accomplished by many methods; we investigate rerandomization and a pairwise greedy-switching heuristic (the subject of our previous work) where this heuristic now operates on the differences of the covariate values within each match in contrast to the raw subject covariate values. 

To illustrate the advantages of our hybrid experimental design that both matches and imbalance-optimizes, consider linear and nonlinear response models, respectively $y_i = x_i + \half\betaT w_i + \errorrv_i$ and $y = x^2 + \half \betaT w_i + \errorrv$ where $\betaT$ we set to 1, $x_i$'s are drawn iid from $U(0, 3)$ and the $\errorrv_i$'s are drawn iid from $\normnot{0}{0.1^2}$. We use a small sample size of $2n = 40$ and consider three designs: (G) pure imbalance optimization via the greedy pair switching design which will be detailed in Section~\ref{sec:methodology}, (M) optimal non-bipartite matching and (MG) optimal non-bipartite matching followed by imbalance optimization via the greedy pair switching design. We simulate the above system 1,000 times and measure covariate imbalance using the metric log$_{10}(|\xbarT - \xbarC|)$ and squared error between the estimate $\ybarT - \ybarC$ and $\betaT$. We average over the simulations to estimate both means and display the results in Table~\ref{tab:basic_demo}.

\begin{table}[h]
\centering
\begin{tabular}{cccc|ccc}
&&Average & Average & \multicolumn{3}{c}{p value for the average} \\
Response && Log$_{10}$  & Squared & \multicolumn{3}{c}{squared error comparison} \\
 Model & Design & Imbalance  &  Error & G & M & MG \\ \hline
Linear & G  & -3.90 & 0.00099 & - & $\approx 0$ & 0.74  \\
Linear & M  & -1.87 & 0.00135 & - & - & $\approx 0$  \\
Linear & MG  &  -4.55 & 0.00105 & - & - & - \\ \hline
Nonlinear & G &  - &0.04350 & - & $\approx 0$  & $\approx 0$ \\
Nonlinear & M &  - &0.00614 & - & - & 0.08\\
Nonlinear & MG &  - & 0.00273 & - & - & - \\
\end{tabular}
\caption{Results of the illustrative simulation explained in the text.}
\label{tab:basic_demo}
\end{table}

In the linear model, the pure imbalance-optimizing design (G) provides better MSE performance than matching (M) and employing the hybrid design combining matching with imbalance-optimization (MG) does not penalize performance (i.e. G and MG have MSE's that are statistically equal). When the model is non-linear, the covariate imbalance is less relevant, and  thus (M) provides much better MSE performance than (G) and the hybrid design (MG) outperforms (M) although not statistically significantly here (we will see in the results section that this advantage is likely only for very small sample sizes). Thus the hybrid (MG) design is robust and adaptive as it has the best performance in both linear and nonlinear settings.

When examining the imbalance, M achieves small covariate imbalance but not when compared to G / MG which is expected as that is not the former's primary objective in contrast to the latters' primary objective. MG has smaller covariate imbalance than G as it can combine the reduction in imbalance from M with the largely independent reduction in imbalance from G, the pure imbalance-optimizing procedure. The nearly order of magnitude difference in imbalance between MG and G does not translate to a significantly lower MSE as there is diminishing returns for imbalance optimization; the theoretical reasons for this will become clear in the following section.


\section{Methodology}\label{sec:methodology}

\subsection{Setup and Assumptions}\label{subsec:setup}

We denote the responses $\y = \bracks{y_1, \ldots, y_{2n}}^\top$ where the number of subjects $n$ is assumed even. The \emph{assignment} or \emph{allocation} vector is $\w = \bracks{w_1, \ldots, w_n}^\top$ whose entries are either +1 (the subject received $T$) or -1 (the subject received a $C$) and $\w \in \braces{-1,+1}^{2n}$. We define a \emph{design} $D$ as $W_D$, a discrete uniform random variable with support $\allocspace_D \subseteq \braces{-1,+1}^{2n}$ and number of vectors $\abss{\allocspace_D} \leq 2^{2n}$. We restrict the designs we consider to those that have the mirror property \citet[Assumption 2.2]{Kapelner2020}, \citet[Assumption 1b]{Kallus2018}, \citet{Nordin2020} i.e. for any $\w$, if treatment assignments and control assignments were switched then that resulting assignment would also be in the design: $\w \in \allocspace \Rightarrow \onevec - \w \in \allocspace$. We further restrict the designs we consider to be \emph{forced balance procedures} where all allocations have the same number of treated and control subjects \citep[Chapter 3.3]{Rosenberger2016}. This is a minor restriction is a standard design known as the balanced complete randomization design (BCRD) or completely random design, denoted $\allocspace_{BCRD} = \braces{\w\,:\, \sum w_i = 0}$ which has $\binom{2n}{n}$ assignments. If a design further restricted this set, i.e. $\allocspace_D \subset \allocspace_{BCRD}$, we call $D$ a \emph{restricted design}.

We assume the only source of the randomness in the response is thus the treatment assignments $\w$. This assumption on the source of randomness is termed the \emph{randomization model} \citep[Chapter 6.3]{Rosenberger2016}, the \textit{Fisher model} or the \textit{Neyman model} whereby \qu{the $2n$ subjects are the population of interest; they are not assumed to be randomly drawn from a superpopulation} \citep[page 297]{Lin2013}.

As in \citet{Kallus2018}, we employ the simple \emph{differences-in-means} estimator, 

\bneqn\label{eq:estimator}
\betaThat := \frac{\w^\top \y}{2n} = \half(\YbarT - \YbarC)
\eneqn 

which is population average treatment estimator regardless of the response model and the effect of the treatment \citep[Example 4.7]{Lehmann1998}.

\subsection{Previous Literature}

Our design idea stems from a combination of two long-standing streams of research: first, the pure-imbalance optimizing design literature and second, the binary-matching bias-reducing literature.

For the pure-imbalance optimizing design, we first reference the rerandomization design R. Here, one begins with vectors from BCRD and discards those whose covariate imbalance is beyond a desired threshold. This idea that goes back to Fisher, the father of randomized experiments who was aware of imbalanced allocations and ironically warned against pure unrestricted randomization, the procedure that he is famous for. 

Here, $\allocspace_{R(a)} := \braces{\w\,:\, d_M(\w) \leq a, \w \in \allocspace_{BCRD}}$ where 

\bneqn\label{eq:mahal}
d_M(\w) := n (\xbarT - \xbarC)^\top \hat{\Sigma}_X^{-1} (\xbarT - \xbarC),
\eneqn

a popular metric of covariate imbalance among the treatment and control groups known as the Mahalanobis distance, $a$ is an upper-bound acceptability threshold of the covariate imbalance, $\xbarT$ and $\xbarC$ are the averages across the $p$ covariates in the treatment and control groups respectively and $\hat{\Sigma}_X$ is the sample variance-covariance matrix of all $2n$ subjects. Since $a < \infty$ implies that $\allocspace_{R(a)} \subset \allocspace_{BCRD}$, rerandomization is a restricted design by construction. Assuming multivariate-normally distributed covariates and an additive treatment effect but no assumption on the nature of the response model, the multiplicative reduction in MSE of the difference-in-averages treatment estimator when using the R design is $\eta(p, a) R^2$ where 

\bneqn\label{eq:rerand_reduction}
\eta(p, a) := 1 - \frac{2}{p}  \frac{\gamma(p/2 + 1, a/2)}{\gamma(p/2, a/2)},
\eneqn 

$\gamma(\cdot, \cdot)$ denotes the lower incomplete gamma function and $R^2$ is variance explained in an OLS linear regression of the response on the covariates \citep{Morgan2012}.

This reduction in treatment estimator error comes from two sources (1) a reduction of the imbalance in the $p$ covariates which is at their control and (2) the degree of linearity of the covariates in the response as measured by $R^2$. 

As for their first source of estimator error reduction, lowering the imbalance in rerandomization is only slightly more impressive than BCRD. There are many ideas for achieving better reduction in covariate imbalance. First, \citet[Section 3.3]{Kallus2018} conjectured using a heuristic argument that the optimal reduction (i.e. for the best possible vector) is exponential $2^{-\Omega(n)}$. Numerical optimization such as in \citet{Bertsimas2015} employ heuristics to approximate the optimal vector. Some heuristics come with theoretical guarantees, e.g. the greedy pair-switching design of \citet{Krieger2019} (henceforth denoted KAK19) uses a greedy heuristic that begins with a draw from BCRD, considers switching every T/C pair and retains the one with the greatest reduction in the imbalance metric and stops switching when it is no longer fruitful. The resulting vectors have a provably very low imbalance $O_p(n^{-(1 + 2/p)})$ which can then be further enhanced by generating many $\w$'s and retaining only the best of those akin to rerandomization. Moreover, since there is a provably low number of switches, the degree of randomness is nearly that of BCRD. We denote this restricted design as G.

Better rates than offered by G are unlikely to matter as the imbalance reduction multiple (Equation~\ref{eq:rerand_reduction}) has diminishing returns in $a$. After a procedure such as KAK19, the performance of the estimator is much more sensitive to the second source of error, i.e. the strength of the covariates' affect on the response (as gauged by $R^2$). If the $R^2$ is low (in the case of a non-linear response model and/or a high noise setting), covariate imbalance reduction will not be too fruitful. Moreover, \citet[Section 2.3.3]{Kallus2018} proves that minimizing $a$ in rerandomization is fruitful for minimax estimator error reduction only in the case of a linear response model. \citet{Kallus2018} explains the reason why this advantage is only limited to linear response models: even if $d_M(\w) = 0$ (i.e. the first moments are matched perfectly), the covariate distributions in the two arms could be very different. In a nonlinear response model, the estimator performance can depend on more subtle differences in the covariates' distributions (e.g. their tail behavior).

The second stream of research, binary matching, implicitly attempts to equalize the covariate distributions in both arms. Here, the $2n$ subjects' indices are first organized into a \textit{matched pair structure} $\mathcal{M}_d$, a set whose elements are \textit{pairs} (a set of two subject indices). The structure $\mathcal{M}_d$ is created by first assuming a distance function $d(\x_r, \x_s) : \reals^p \times \reals^p \rightarrow \reals_{\geq 0}$ whose inputs are subject covariates. A $2n \times 2n$ nonnegative symmetric matrix is then computed consisting of distances for all pairs of subjects. The optimal match structure $\mathcal{M}_d^\star$ would be one of the $\binom{2n}{n}$ sets that minimizes the sum of the resulting intramatch distances i.e. $\sum_{\braces{r,s} \in \mathcal{M}} d(\x_r, \x_s)$. The algorithm that produces the optimal match structure is known as \emph{optimal nonbipartite matching}. This algorithm has been shown to be reducible to a polynomial-time algorithm (see \citealp{Lu2011} for details and history). To create an assignment vector $\w$, the individual assignments within each of the $n$ pairs are randomly allocated to T/C or C/T via $n$ iid Bernoulli draws. We denote the optimal binary matching design with the Mahalanobis distance metric (Equation~\ref{eq:mahal}) as M and its allocation space by $\allocspace_{M} = \{\w\,:\,w_r = -w_s, \braces{r,s} \in \mathcal{M}_{d_M}^\star\} \subset \allocspace_{BCRD}$. 

Can binary matching play a role in experimental design? This M design was investigated by \citet{Greevy2004} who reported better imbalance and higher power using randomization tests. However, matching in order to have robust estimation under a nonlinear response model has a long literature in observational studies and we recommend \citet{Stuart2010} for a broad overview. (Without the ability to manipulate $\w$, the matching procedure is quite different as it is limited to fixed $w_i$'s and is called \emph{bipartite matching}). \citet{Rubin1979} was the first to show this robustness and recommended that $d$ be specified as the Mahalanobis distance of Equation~\ref{eq:mahal}. 

Further, \citet[Section 2.3.2]{Kallus2018} proves that binary matching via $d$ is the minimax variance minimizing experimental design for response functions that are Lipschitz continuous with respect to metric $d$. It is reasonable to conclude in a real-world experimental setting (for instance in a clinical trial in medicine) that the response would be continuous with derivatives not changing too quickly (i.e. satisfying the Lipschitz conditions). Thus, to reduce the variance of the treatment estimator in nonlinear response models, we elect to use a binary matching design.

Our contribution is relatively simple: we first compute $\mathcal{M}_{d_M}^\star$ as part of an M design and this binary pairing structure will provide robustness to nonlinear response models. Then instead of assigning the specific pairs' subjects via Bernoulli draws as explained above, we further restrict the pair assignments to insist on small imbalance by using R (resulting in the matching-then-rerandomization design we term MR) or by using G (resulting the matching-then-greedy-pair-switching design we term MG).  We turn to the details of MR and MG now.

\subsection{Our Algorithms}

First, $\mathcal{M}_{d_M}^\star$ is computed by running the optimal non-bipartite matching algorithm implemented by \citet{Beck2016} in the \texttt{R} package \texttt{nbpMatching}.

For R after M, $\allocspace_{MR(a)} := \braces{\w\,:\, d_M(\w) \leq a, \w \in \allocspace_M}$ i.e. we draw many different assignments from $\allocspace_M$ and retain those whose Mahalanobis distances meet the threshold $a$ thereby finding small imbalances subject to the match structure.

For G after M, we cannot simply input $\w \in \allocspace_M$ directly into KAK19 as this would violate the $\mathcal{M}_{d_M}^\star$ structure, breaking the binary pairings. Instead, we temporarily view the matched pairs as individual subjects (of which there are $n$) as the input into KAK19. This means we temporarily view the intramatch covariate vector differences as the subjects' covariates which are inputted into KAK19. Put another way, the T arm now refers to a matched pair assigned T/C where the covariate difference vector is computed by taking the first subject's covariates minus the second subject's covariates. The C arm would be the opposite. In one iteration of KAK19, all \textit{pair of pairs switches} are considered: from (T/C)/(C/T) to (C/T)/(T/C) or vice versa. One of these will result in the largest reduction in imbalance as measured among the pair of pairs which is algebraically the same imbalance as measured among the entire $\w$ vector for the $2n$ subjects. This best pair of pairs switch is retained and the algorithm stops when an iteration cannot reduce the imbalance any longer. This scheme thereby preserves $\mathcal{M}_{d_M}^\star$.

Under our designs, to evaluate hypotheses concerning the value of $\beta_T$ we can use the randomization test and to construct approximate confidence intervals for $\beta_T$ we can invert the randomization test. For details, see \citet[Section 2.2]{Morgan2012} and KAK19, Section 6; both provide a careful step-by-step explanation and a literature review.

One can consider an alternative implementation of the G in MG as follows. After M, each iteration of G can perform one switch of one pair (instead of a switch of a pair of pairs). Here, we now have fewer switch possibilities, i.e. $n$ and not $\binom{n}{2}$, and we conjecture this procedure would not perform as well as the one proposed above. 

One can also consider a GM (or RM) procedure i.e. to first imbalance-optimize via G (or R) and then employ binary matching. In that scenario, there will be a different $\mathcal{M}$ for each initial imbalance optimized vector based on a bipartite matching. The gain here would be that the imbalance-optimization would be slightly better but it comes at a cost: bipartite matches are much more constrained than nonbipartite matches and thus we conjecture that these designs will have much lower robustness to nonlinear models. 

\subsection{Theoretical Investigation of MG and MR}\label{sec:theory}

\subsubsection{Imbalance}

The key idea of our designs is to retain the match structure while optimizing imbalance. Instead of employing G or R using $\allocspace_{BCRD}$ as a base, we now employ G or R using $\allocspace_{M}$ as a base. We are not aware of theoretical results for the imbalance after optimal nonbipartite matching when $p > 1$ and thus we focus on the case when $p = 1$. 

In this univariate case, consider the order statistics $X_{(1)}, \ldots, X_{(2n)}$. Optimal matching creates $\mathcal{M} = \braces{\braces{2n, 2n-1}, \ldots, \braces{2,1}}$, i.e. pairing the first largest covariate with the second largest covariate value, then pairing the third largest with the fourth largest, etc. We will consider the case where $X_1,\ldots,X_{2n}$ are iid $\stduniform$. Define $\Delta_i = X_{(2i)} - X_{(2i-1)}$ for $i = 1, 2, \ldots, n$. By \citet[Theorem 6.6c]{Dasgupta2011}, $\bracks{\Delta_1,\ldots,\Delta_n}$ has the same distribution as $\oneover{S}\bracks{E_1,\ldots,E_n}$, where $E_1,\ldots,E_n,\ldots,E_{2n+1}$ are iid $\exponential{1}$ and $S := \sum_{i=1}^{2n+1} E_i = O_p(n)$. Since the denominator $S$ is common to all entries of the vector, applying the greedy pair-switching of KAK19 to $\bracks{\Delta_1,\ldots,\Delta_n}$ is the same as (in terms of distribution) applying the algorithm to $\bracks{E_1,\ldots,E_n}$ and then dividing by $S$. 

The imbalance of assignments in G is $O_p(n^{-3})$ by KAK19 (Theorem 1), a distribution-free result. When dividing by $S$, which is $O_p(n)$, the resultant rate of our MG design is $O_p(n^{-4})$.


The imbalance of assignments in R is less than $a$ by construction. However, if $a$ is too small, then practical computational constraints dictate that all BCRD assignments one generates will have imbalance greater than $a$, resulting in the inability to locate any assignments in $\allocspace_{R(a)}$. KAK19 reframes the discussion of R's imbalance by making this computational constraint explicit. Let $n_R$ denote the number of BCRD assignments one searches through for each resultant assignment in R. For every collection of $n_R$, the assignment that provides the minimum imbalance in that collection is retained. These retained vectors then have imbalance which is order $n_R^{-1} O_p(n^{-1/2})$ where $O_p(n^{-1/2})$ is the imbalance of BCRD. It then follows by the same argument above that when dividing by $S$, the resultant imbalance of MR is then $n_R^{-1} O_p(n^{-3/2})$. 

Extending the theoretical results for imbalance in MG beyond the assumption of uniformly distributed covariates is difficult. The main result needed to prove that the order of the imbalance of the assignments in G is $O_p(n^{-3})$ is to show that there exists a pair of covariate values to switch $X_i, X_j$ where $w_i = 1$ and $w_j = -1$ that result in a necessary reduction in imbalance. The analogous result here would require the locating of $\Delta_i, \Delta_j$ where $w_i = 1$, $w_{i+1} = -1$, $w_j = -1$ and $w_{j+1} = 1$ that results in the same reduction in imbalance after switching both pairs. Showing the necessary reduction in imbalance assumed that the $X_i$'s are continous and iid. We were able to use this result in this paper when the $X_i$'s were uniformly distributed, but if the $X_i$'s are normally distributed, it is much more complicated as the $\Delta_i$'s corresponding to covariate values in a tail are stochastically larger than the $\Delta_i$'s corresponding to covariate values in the center. The imbalance of MG in the normal case is shown by simulation to be lower than G (see Figure~\ref{fig:mse_density_plots_normal} in the Supporting Information).

\subsubsection{Degree of Randomness}

We now turn to degree of randomness in MG and MR. We reiterate that in experimental design, it is important for the assignments to be highly random as this randomness provides insurance against a large variance in the treatment effect estimator due to unobserved covariates \citep{Kapelner2020}. Both G and R are less random than BCRD since they are forms of restricted randomization. In the case of G, KAK19 (Theorem 2 and Proposition 1) shows that G's assignments are asymptotically as random as BCRD according to three degree-of-randomness metrics: (a) the pairwise entropy metric of assignments, (b) the standard error metric of probability of pairwise assignments and (c) the accidental bias metric of \citet[Section 5]{Efron1971} defined as the maximum eigenvalue of the variance covariance matrix of $W$, the random variable producing assignments in any strategy (see Section~\ref{subsec:setup}). The reason for G's high degree of randomness is that the number of pairwise switches made in G is small, being only $O_p(\sqrt{n})$. Thus, since G starts with BCRD, and its assignments are not changed significantly, the resulting $\allocspace_G$ is very large. In the case of R, we directly limit the number of designs based on the cutoff $a$. As, $a$ decreases, $\allocspace_R$ shrinks relative to $\allocspace_{BCRD}$, making R less random. Since $a$ can never be too small due to computational limitations, it follows that randomness is not substantially deteriorated in R. Thus, MG is as random as M asymptotically and MR-in-practice is likely as random as M asymptotically as well. And M is highly random according to all three degree-of-randomness metrics, only slightly less random than BCRD. 

\section{Simulation Results}\label{sec:simulations}

To illustrate our main results we turn to simulations. All simulations herein were performed with \texttt{GreedyExperimentalDesign}, an \texttt{R} package available on \texttt{CRAN} whose core is implemented in \texttt{Java} for speed. We measure mean squared error (MSE) of the difference-in-averages estimator for an additive average treatment effect when the response model depends on underlying covariate(s). 

We conjecture that (a) the MSE performance of MR/MG will be similar to imbalance-optimizing designs G/R when the response model is purely linear, (b) the MSE performance of MR/MG will be similar to the pure nonbipartite matching design M when the response model is purely nonlinear and (c) the MSE performance of MR/MG will be better than either exclusively imbalance-optimizing G/R or exclusively nonbipartite matching M when the response model is a hybrid of a strong linear component and strong nonlinear component.

To demonstrate the conjecture we consider $n=100$, two covariates and five response models that take the form $\Y = \betaT\w + f(x_1, x_2) + \berrorrv$ where $x_1, x_2$ are independent draws from $U(-\sqrt{3},+\sqrt{3})$ so that their variance is one and the $n$ components of $\berrorrv$ are iid draws from $\normnot{0}{0.5^2}$. The five models are (Z) a null zero model $f = 0$ for calibration and ensuring the simulation is working (L) a purely linear model $f = 3x_1 + 3x_2$, (LsNL) a mostly linear but slightly nonlinear model $f = 3x_1 + 3x_2 + x_1^2$, (LNL) a linear and nonlinear model $f = 3x_1 + 3x_2 + x_1^2 + x_2^2 + x_1 x_2$ and (NL) a purely nonlinear model $f = x_1^2 + x_2^2 + x_1 x_2$. 

We then consider the six designs of the previous section all of which feature an equiprobable draw from a set of mirrored allocations: (BCRD) balanced and completely randomized, (R) rerandomization of \citet{Morgan2012} where only the top 1\% of BCRD vectors are retained, (G) the greedy pair switching of KAK19, (M) optimal nonbipartite matching, and our two new hybrid approaches of (MG) optimal nonbipartite matching followed by greedy pair switching on the within-pair covariate differences and (MR) optimal nonbipartite matching followed by rerandomization on the within-pair covariate differences where the top 1\% of pairings are retained. Thus, we have $5 \times 6$ = 30 sample size-model-design simulation settings.

For each sample size-model-design simulation setting, we simulate 50 different covariate and noise value settings, $x_1, x_2, \errorrv_1, \ldots, \errorrv_n$. Within each specific covariate-noise realization setting, we draw 500 unique allocation vectors from each of the six design strategies. We define MSE as $\expesubnostr{X_1, X_2, \berrorrv}{\cexpesubnostr{W_D}{(\hat{\beta}_T - \betaT)^2}{X_1, X_2}}$ for each sample size-model-design setting. This MSE is estimated by first taking an arithmetic average of the 500 squared errors within the covariate-noise setting and then averaging these over the 50 covariate-noise settings. Estimates for matching designs (M and MR) usually are computed by the average paired differences. This estimate is algebraically equivalent to the naive difference-in-averages estimate and thus estimates are computed uniformly for all six designs.

\begin{figure}[h]
\centering
\includegraphics[width=6.5in]{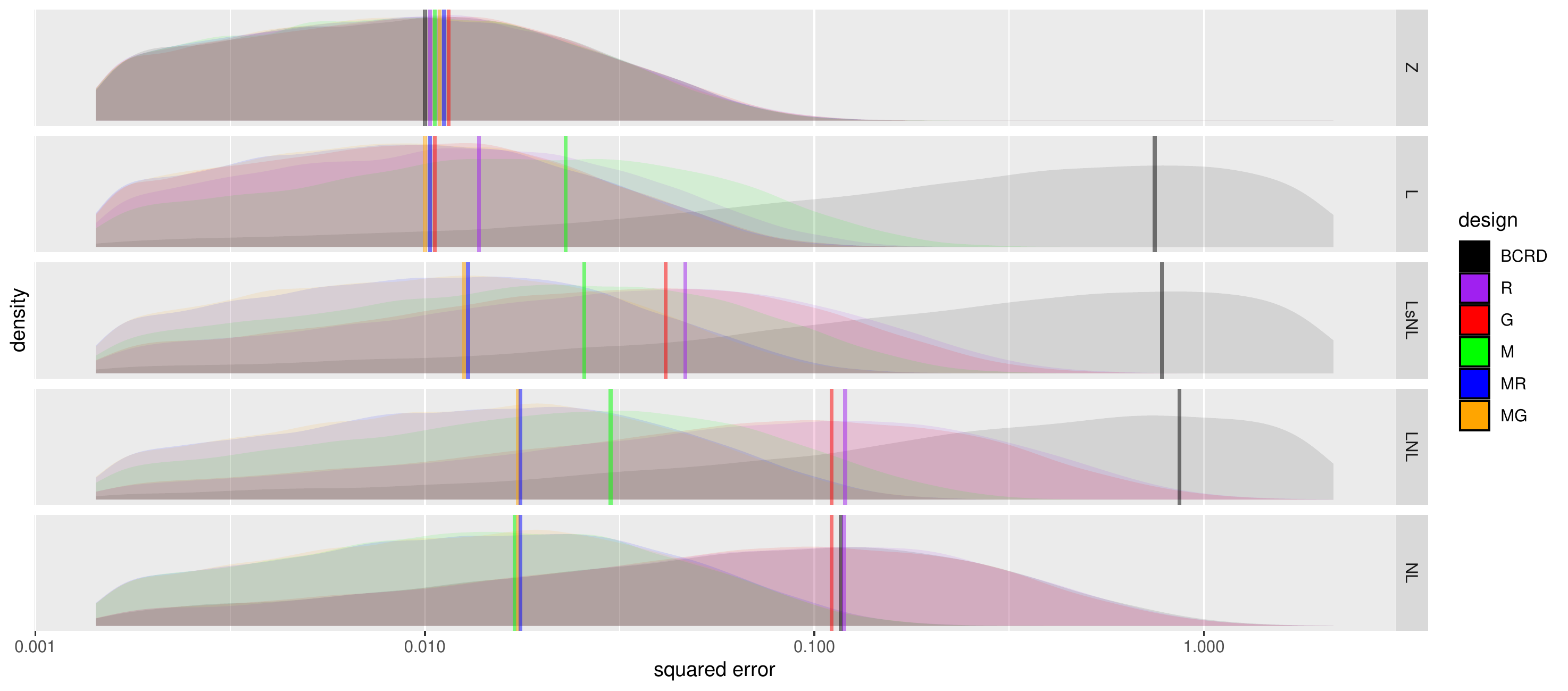}
\caption{Density plots for MSE (on a log scale) over the covariate distributions, noise and assignments when $n=100$ for the five models (the vertically stacked panes) and six designs (depicted as different colors) explained in the text. The vertical lines indicate the MSE estimate (explained in the text) by design. Lines directly adjacent to one another are the result of a slight jitter. Statistically significant differences are tabulated in Table~\ref{tab:all_pvals}.}
\label{fig:mse_density_plots}
\end{figure}

Figure~\ref{fig:mse_density_plots} provides density plots for MSE when $n=100$ for all five models and six designs and Table~\ref{tab:all_pvals} tabulates the statistically significance of each design comparison. The null response model (Z) serves as a check on the integrity of our simulation: since the covariates are disjointed from the response, all designs perform equally as seen in the top pane of Figure~\ref{fig:mse_density_plots} and for this reason it was omitted from Table~\ref{tab:all_pvals}. For all other response models where the covariates inform the response, the figure and tables illustrates the main thrust of this paper's contribution: (1) Our designs, MR and MG, are always the best performers (or tied for the best performance) indicating robustness to linear, nonlinear and mixtures of linear and nonlinear response models. Other observations support this main thrust. (2) BCRD is the worst performer (or tied for worst performance) which is expected given that it does not allocate subjects with the benefit of knowing the covariate values. (3) In the exclusively linear model (L), the pure imbalance optimizing designs perform the best i.e. G, R, MR, MG all perform equally with M lagging. (4) In the exclusively nonlinear model (NL), the matching designs perform the best i.e. M, MR, MG all perform equally. (5) In the linear-nonlinear models (LsNL, LNL), our matching-plus-imbalance-optimizing designs MR and MG are equally the best, the matching M is second best and the pure imbalance optimizing designs G and R lag behind M. (6) R lags behind G because its imbalance is a much smaller rate (see KAK19 for details why). MR also lags behind MG for the same reason but the difference is not appreciable in these simulations.

\begin{table}[h]
\centering

\begin{subtable}{.5\linewidth}{
\begin{tabular}{rlllll}
 & R & G & M & MR & MG \\ 
  \hline
BCRD & *** & *** & *** & *** & *** \\ 
  R &  &  &  &  &  \\ 
  G &  &  & ** &  &  \\ 
  M &  &  &  & ** & ** \\ 
  MR &  &  &  &  &  \\ 
\end{tabular}
}
\caption{Model: L}
\label{tab:pvals1}
\end{subtable}~
\begin{subtable}{.5\linewidth}{
\begin{tabular}{rlllll}
 & R & G & M & MR & MG \\ 
  \hline
BCRD & *** & *** & *** & *** & *** \\ 
  R &  &  & *** & *** & *** \\ 
  G &  &  & ** & *** & *** \\ 
  M &  &  &  & * & * \\ 
  MR &  &  &  &  &  \\ 
\end{tabular}
}
\caption{Model: LsNL}
\label{tab:pvals2}
\end{subtable}\\
~\\~\\

\begin{subtable}{.5\linewidth}{
\begin{tabular}{rlllll}
 & R & G & M & MR & MG \\ 
  \hline
BCRD & *** & *** & *** & *** & *** \\ 
  R &  &  & *** & *** & *** \\ 
  G &  &  & *** & *** & *** \\ 
  M &  &  &  & * & * \\ 
  MR &  &  &  &  &  \\ 
\end{tabular}
}
\caption{Model: LNL}
\label{tab:pvals3}
\end{subtable}~
\begin{subtable}{.5\linewidth}{
\begin{tabular}{rlllll}
 & R & G & M & MR & MG \\ 
  \hline
BCRD &  & *** & *** & *** & *** \\ 
  R &  & *** & *** & *** & *** \\ 
  G &  &  & *** & *** & *** \\ 
  M &  &  &  &  &  \\ 
  MR &  &  &  &  &  \\ 
\end{tabular}
}
\caption{Model: NL}
\label{tab:pvals4}
\end{subtable}

\caption{Tukey-Kramer comparisons of the MSE for each design pair by model. * indicates $p < 0.05$, ** indicates $p < 0.01$ and *** indicates $p < 0.001$. The Z model is not displayed as none of its design pairs' MSEs tested statistically significant.}
\label{tab:all_pvals}
\end{table}

We also repeated the simulation for three other sample sizes ($n=32$, $n=132$ and $n=200$) and they all have qualitatively similar results. The tabulation of MSE estimates for all sample sizes, designs and response models can be found in Tables~\ref{fig:ols_n_32}-\ref{fig:ols_n_200} in the Supporting Information. The entire simulation was also repeated for the case of normally distributed covariates and the results were not substantially different (see Figure~\ref{fig:mse_density_plots_normal} in the Supporting Information).

\section{Concluding Remarks}\label{sec:conclusion}

We proposed a randomized experimental design that is a hybrid between two well studied strategies: those that optimize covariate imbalance and those that use binary matching. The former is known to be minimax optimal when the response model is linear in the covariates and the latter is known to be minimax optimal when the response model is continuous with limits on its derivative with respect to all $p$ covariates. Additionally, we have theory that shows that the fusing of both strategies fortuitously enhances the covariate imbalance all while retaining a degree of randomization close to that of the classic completely random designs. 

Because we fuse both types of design together without sacrificing the performance of either, our designs provides very low MSE performance when estimating a population average treatment effect in the purely linear case (because we optimize covariate imbalance), the purely nonlinear case (as we our assignments are optimal nonbipartite matched pairs) and response models that feature both linear and nonlinear components. Thus we expect our design to be very powerful when employed in real-world settings such as clinical trials and Internet-based experimentation.

There are many extensions to this work. First, there are different types of matching such as ratio matching \citep[Section 3.1.2]{Stuart2010}, matching in more than two arms (ibid, Section 6.1.4), matching with discrete covariates (the Mahalanobis distance function is not the most appropriate here as discussed in \citealt{Gu1993}) and matching with some covariates being more important than others. As some matches may not be acceptable, one can caliper match; this would result in a subset of the subjects being matched and a subset being unmatched. In this case, the G algorithm would have to be redesigned but the R procedure would be fairly straightforward. Further, one may wish to use the OLS estimator instead of the classical difference-in-averages estimator (Equation~\ref{eq:estimator}). One can argue that restricted randomization can leave you vulnerable to high MSE estimation due to imbalance in unobserved covariates \citep[Sections 2.2.1, 2.2.3, 2.3.1 and 2.3.3]{Kapelner2020}. Therein, they proved that restricted designs with a high degree of randomness, such as our hybrid designs in this work (see Section~\ref{sec:theory}), are highly random and thus are not susceptible. Further, \citet[Equation 14]{Kapelner2020} showed that the OLS estimator improves asymptotic MSE by an order of magnitude in sample size. However, imbalance-optimizing designs (such as R and G) still provide improvement in the linear case. Covariate adjustment cannot aid with the nonlinear component of the response model and thus our hybrid designs will perform very well in conjunction with the OLS estimator on nonlinear models. And the OLS estimator is actually preferred in matching designs \citep{Rubin1979}. Future work can also adapt this design to the sequential setting, where individuals in the experiment arrive one by one and must be assigned quickly to an arm. We can apply the on-the-fly matching procedure in \citet{Kapelner2014} with the rerandomization approach in \citet{Zhou2018} in the same vein as our hybrid procedure. 

\subsection*{Replication}

All figures and tables can be reproduced by running the \texttt{R} code found at \url{https://github.com/kapelner/GreedyExperimentalDesign/blob/master/hybrid_paper}.

\subsection*{Acknowledgements}

This research was supported by Grant No 2018112 from the United States-Israel Binational Science Foundation (BSF).

\bibliographystyle{apa}\bibliography{refs}

\pagebreak
\appendix

\section*{Supporting Information}\label{app}
\renewcommand{\theequation}{A\arabic{equation}}
\renewcommand{\thefigure}{A\arabic{figure}}
\renewcommand{\thetable}{A\arabic{table}}
\setcounter{figure}{0}  
\setcounter{table}{0}

\begin{figure}[h]
\centering
\includegraphics[width=6.5in]{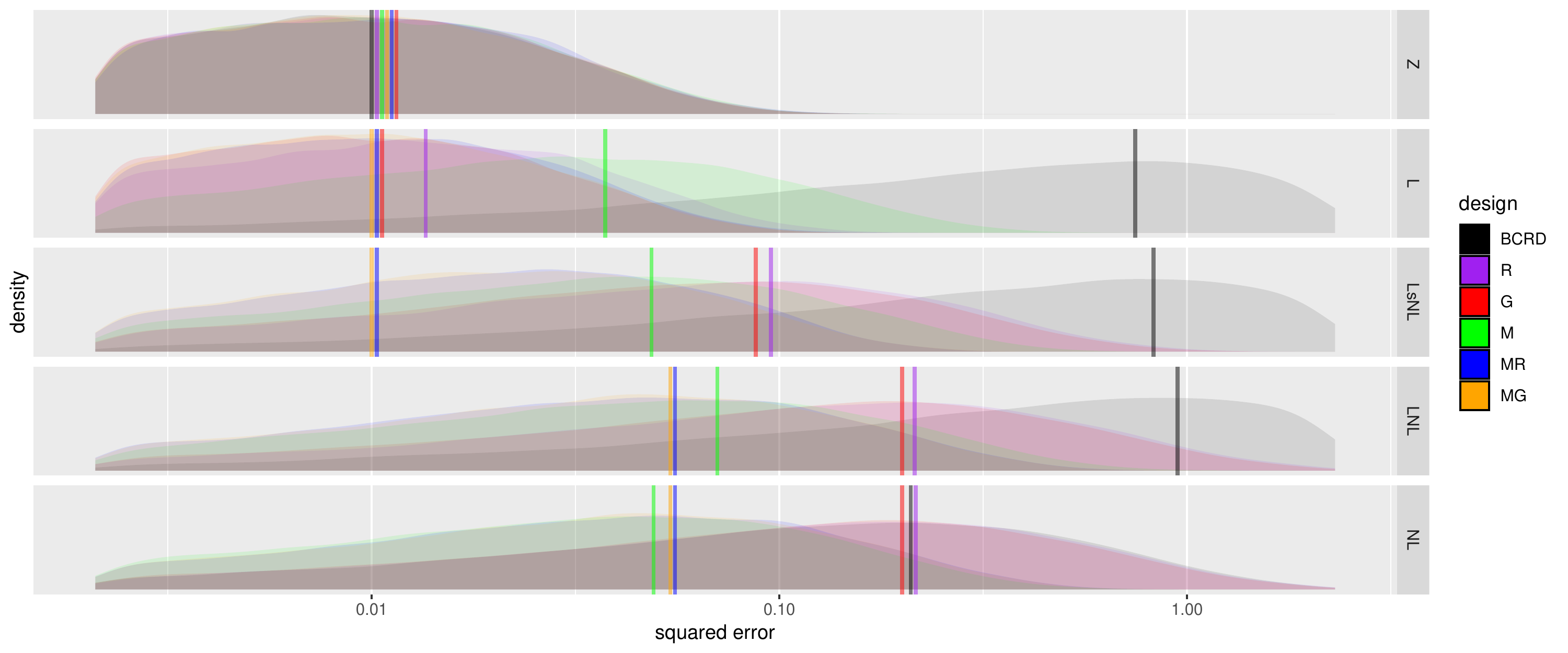}
\caption{The same density plots as Figure~\ref{fig:mse_density_plots} except that the underlying simulation used normally distributed covariates.}
\label{fig:mse_density_plots_normal}
\end{figure}

\begin{table}[ht]
\centering
\begin{tabular}{rrrrr}
  \hline
 & Estimate & Std. Error & t value & Pr($>$$|$t$|$) \\ 
  \hline
(Intercept) & 0.0100 & 0.0022 & 4.57 & 0.0000 \\ 
  M & -0.0000 & 0.0031 & -0.01 & 0.9917 \\ 
  R & -0.0000 & 0.0031 & -0.00 & 0.9999 \\ 
  G & -0.0001 & 0.0031 & -0.04 & 0.9659 \\ 
  MR & -0.0001 & 0.0031 & -0.02 & 0.9809 \\ 
  MG & -0.0001 & 0.0031 & -0.02 & 0.9806 \\ 
  L & 0.7016 & 0.0031 & 226.58 & 0.0000 \\ 
  LsNL & 0.7284 & 0.0031 & 235.22 & 0.0000 \\ 
  LNL & 0.8049 & 0.0031 & 259.91 & 0.0000 \\ 
  NL & 0.1040 & 0.0031 & 33.59 & 0.0000 \\ 
  M:L & -0.6887 & 0.0044 & -157.25 & 0.0000 \\ 
  R:L & -0.6979 & 0.0044 & -159.37 & 0.0000 \\ 
  G:L & -0.7016 & 0.0044 & -160.22 & 0.0000 \\ 
  MR:L & -0.7016 & 0.0044 & -160.20 & 0.0000 \\ 
  MG:L & -0.7016 & 0.0044 & -160.22 & 0.0000 \\ 
  M:LsNL & -0.7131 & 0.0044 & -162.84 & 0.0000 \\ 
  R:LsNL & -0.6918 & 0.0044 & -157.98 & 0.0000 \\ 
  G:LsNL & -0.6970 & 0.0044 & -159.16 & 0.0000 \\ 
  MR:LsNL & -0.7257 & 0.0044 & -165.71 & 0.0000 \\ 
  MG:LsNL & -0.7259 & 0.0044 & -165.76 & 0.0000 \\ 
  M:LNL & -0.7842 & 0.0044 & -179.07 & 0.0000 \\ 
  R:LNL & -0.6965 & 0.0044 & -159.04 & 0.0000 \\ 
  G:LNL & -0.7060 & 0.0044 & -161.20 & 0.0000 \\ 
  MR:LNL & -0.7970 & 0.0044 & -181.98 & 0.0000 \\ 
  MG:LNL & -0.7972 & 0.0044 & -182.03 & 0.0000 \\ 
  M:NL & -0.0966 & 0.0044 & -22.05 & 0.0000 \\ 
  R:NL & 0.0005 & 0.0044 & 0.11 & 0.9161 \\ 
  G:NL & -0.0051 & 0.0044 & -1.17 & 0.2431 \\ 
  MR:NL & -0.0962 & 0.0044 & -21.96 & 0.0000 \\ 
  MG:NL & -0.0963 & 0.0044 & -21.99 & 0.0000 \\ 
   \hline
\end{tabular}
\caption{OLS output for squared error regressed on design $\times$ response model in the simulation of Section~\ref{sec:simulations} for $n = 32$.}
\label{fig:ols_n_32}
\end{table}

\begin{table}[ht]
\centering
\begin{tabular}{rrrrr}
  \hline
 & Estimate & Std. Error & t value & Pr($>$$|$t$|$) \\ 
  \hline
(Intercept) & 0.0100 & 0.0022 & 4.57 & 0.0000 \\ 
  M & -0.0000 & 0.0031 & -0.01 & 0.9917 \\ 
  R & -0.0000 & 0.0031 & -0.00 & 0.9999 \\ 
  G & -0.0001 & 0.0031 & -0.04 & 0.9659 \\ 
  MR & -0.0001 & 0.0031 & -0.02 & 0.9809 \\ 
  MG & -0.0001 & 0.0031 & -0.02 & 0.9806 \\ 
  L & 0.7016 & 0.0031 & 226.58 & 0.0000 \\ 
  LsNL & 0.7284 & 0.0031 & 235.22 & 0.0000 \\ 
  LNL & 0.8049 & 0.0031 & 259.91 & 0.0000 \\ 
  NL & 0.1040 & 0.0031 & 33.59 & 0.0000 \\ 
  M:L & -0.6887 & 0.0044 & -157.25 & 0.0000 \\ 
  R:L & -0.6979 & 0.0044 & -159.37 & 0.0000 \\ 
  G:L & -0.7016 & 0.0044 & -160.22 & 0.0000 \\ 
  MR:L & -0.7016 & 0.0044 & -160.20 & 0.0000 \\ 
  MG:L & -0.7016 & 0.0044 & -160.22 & 0.0000 \\ 
  M:LsNL & -0.7131 & 0.0044 & -162.84 & 0.0000 \\ 
  R:LsNL & -0.6918 & 0.0044 & -157.98 & 0.0000 \\ 
  G:LsNL & -0.6970 & 0.0044 & -159.16 & 0.0000 \\ 
  MR:LsNL & -0.7257 & 0.0044 & -165.71 & 0.0000 \\ 
  MG:LsNL & -0.7259 & 0.0044 & -165.76 & 0.0000 \\ 
  M:LNL & -0.7842 & 0.0044 & -179.07 & 0.0000 \\ 
  R:LNL & -0.6965 & 0.0044 & -159.04 & 0.0000 \\ 
  G:LNL & -0.7060 & 0.0044 & -161.20 & 0.0000 \\ 
  MR:LNL & -0.7970 & 0.0044 & -181.98 & 0.0000 \\ 
  MG:LNL & -0.7972 & 0.0044 & -182.03 & 0.0000 \\ 
  M:NL & -0.0966 & 0.0044 & -22.05 & 0.0000 \\ 
  R:NL & 0.0005 & 0.0044 & 0.11 & 0.9161 \\ 
  G:NL & -0.0051 & 0.0044 & -1.17 & 0.2431 \\ 
  MR:NL & -0.0962 & 0.0044 & -21.96 & 0.0000 \\ 
  MG:NL & -0.0963 & 0.0044 & -21.99 & 0.0000 \\ 
   \hline
\end{tabular}
\caption{OLS output for squared error regressed on design $\times$ response model in the simulation of Section~\ref{sec:simulations} for $n = 100$.}
\label{fig:ols_n_100}
\end{table}

\begin{table}[ht]
\centering
\begin{tabular}{rrrrr}
  \hline
 & Estimate & Std. Error & t value & Pr($>$$|$t$|$) \\ 
  \hline
(Intercept) & 0.0073 & 0.0017 & 4.19 & 0.0000 \\ 
  M & 0.0003 & 0.0025 & 0.11 & 0.9126 \\ 
  R & -0.0000 & 0.0025 & -0.02 & 0.9856 \\ 
  G & 0.0000 & 0.0025 & 0.01 & 0.9880 \\ 
  MR & 0.0001 & 0.0025 & 0.05 & 0.9590 \\ 
  MG & 0.0000 & 0.0025 & 0.02 & 0.9841 \\ 
  L & 0.5650 & 0.0025 & 229.80 & 0.0000 \\ 
  LsNL & 0.5907 & 0.0025 & 240.25 & 0.0000 \\ 
  LNL & 0.6414 & 0.0025 & 260.86 & 0.0000 \\ 
  NL & 0.0809 & 0.0025 & 32.92 & 0.0000 \\ 
  M:L & -0.5575 & 0.0035 & -160.34 & 0.0000 \\ 
  R:L & -0.5620 & 0.0035 & -161.64 & 0.0000 \\ 
  G:L & -0.5650 & 0.0035 & -162.50 & 0.0000 \\ 
  MR:L & -0.5650 & 0.0035 & -162.49 & 0.0000 \\ 
  MG:L & -0.5650 & 0.0035 & -162.50 & 0.0000 \\ 
  M:LsNL & -0.5813 & 0.0035 & -167.18 & 0.0000 \\ 
  R:LsNL & -0.5634 & 0.0035 & -162.03 & 0.0000 \\ 
  G:LsNL & -0.5674 & 0.0035 & -163.18 & 0.0000 \\ 
  MR:LsNL & -0.5890 & 0.0035 & -169.39 & 0.0000 \\ 
  MG:LsNL & -0.5890 & 0.0035 & -169.41 & 0.0000 \\ 
  M:LNL & -0.6292 & 0.0035 & -180.95 & 0.0000 \\ 
  R:LNL & -0.5565 & 0.0035 & -160.06 & 0.0000 \\ 
  G:LNL & -0.5644 & 0.0035 & -162.33 & 0.0000 \\ 
  MR:LNL & -0.6369 & 0.0035 & -183.17 & 0.0000 \\ 
  MG:LNL & -0.6368 & 0.0035 & -183.15 & 0.0000 \\ 
  M:NL & -0.0764 & 0.0035 & -21.97 & 0.0000 \\ 
  R:NL & 0.0008 & 0.0035 & 0.22 & 0.8239 \\ 
  G:NL & -0.0040 & 0.0035 & -1.15 & 0.2519 \\ 
  MR:NL & -0.0765 & 0.0035 & -21.99 & 0.0000 \\ 
  MG:NL & -0.0764 & 0.0035 & -21.97 & 0.0000 \\ 
   \hline
\end{tabular}

\caption{OLS output for squared error regressed on design $\times$ response model in the simulation of Section~\ref{sec:simulations} for $n = 132$.}
\label{fig:ols_n_132}
\end{table}

\begin{table}[ht]
\centering
\begin{tabular}{rrrrr}
  \hline
 & Estimate & Std. Error & t value & Pr($>$$|$t$|$) \\ 
  \hline
(Intercept) & 0.0050 & 0.0011 & 4.56 & 0.0000 \\ 
  M & 0.0002 & 0.0015 & 0.11 & 0.9139 \\ 
  R & 0.0000 & 0.0015 & 0.02 & 0.9863 \\ 
  G & 0.0000 & 0.0015 & 0.01 & 0.9901 \\ 
  MR & 0.0002 & 0.0015 & 0.14 & 0.8922 \\ 
  MG & 0.0002 & 0.0015 & 0.15 & 0.8769 \\ 
  L & 0.3512 & 0.0015 & 227.27 & 0.0000 \\ 
  LsNL & 0.3669 & 0.0015 & 237.41 & 0.0000 \\ 
  LNL & 0.4033 & 0.0015 & 260.98 & 0.0000 \\ 
  NL & 0.0506 & 0.0015 & 32.74 & 0.0000 \\ 
  M:L & -0.3482 & 0.0022 & -159.32 & 0.0000 \\ 
  R:L & -0.3493 & 0.0022 & -159.82 & 0.0000 \\ 
  G:L & -0.3512 & 0.0022 & -160.70 & 0.0000 \\ 
  MR:L & -0.3512 & 0.0022 & -160.70 & 0.0000 \\ 
  MG:L & -0.3512 & 0.0022 & -160.70 & 0.0000 \\ 
  M:LsNL & -0.3633 & 0.0022 & -166.20 & 0.0000 \\ 
  R:LsNL & -0.3490 & 0.0022 & -159.66 & 0.0000 \\ 
  G:LsNL & -0.3514 & 0.0022 & -160.76 & 0.0000 \\ 
  MR:LsNL & -0.3662 & 0.0022 & -167.54 & 0.0000 \\ 
  MG:LsNL & -0.3662 & 0.0022 & -167.55 & 0.0000 \\ 
  M:LNL & -0.3987 & 0.0022 & -182.43 & 0.0000 \\ 
  R:LNL & -0.3503 & 0.0022 & -160.25 & 0.0000 \\ 
  G:LNL & -0.3542 & 0.0022 & -162.04 & 0.0000 \\ 
  MR:LNL & -0.4015 & 0.0022 & -183.69 & 0.0000 \\ 
  MG:LNL & -0.4016 & 0.0022 & -183.74 & 0.0000 \\ 
  M:NL & -0.0489 & 0.0022 & -22.37 & 0.0000 \\ 
  R:NL & 0.0005 & 0.0022 & 0.22 & 0.8240 \\ 
  G:NL & -0.0014 & 0.0022 & -0.65 & 0.5173 \\ 
  MR:NL & -0.0487 & 0.0022 & -22.30 & 0.0000 \\ 
  MG:NL & -0.0488 & 0.0022 & -22.34 & 0.0000 \\ 
   \hline
\end{tabular}

\caption{OLS output for squared error regressed on design $\times$ response model in the simulation of Section~\ref{sec:simulations} for $n = 200$.}
\label{fig:ols_n_200}
\end{table}

\end{document}